\documentclass[twocolumn,aps,prb,epsfig,showpacs,superscriptaddress]{revtex4}

\usepackage{graphicx}

\begin{document}

\title{Unravelling the origin of the controversial magnetic properties of BiFeO$_3$ thin films}

\author{H\'el\`ene B\'ea}
\affiliation{Unit\'e Mixte de Physique CNRS/Thales, Route D\'epartementale 128, 91767 Palaiseau, France}

\author{Manuel Bibes}
\email{manuel.bibes@ief.u-psud.fr} \affiliation{Institut d'Electronique Fondamentale, Universit\'e Paris-Sud,
91405 Orsay, France}

\author{Eric Jacquet}
\affiliation{Unit\'e Mixte de Physique CNRS/Thales, Route D\'epartementale 128, 91767 Palaiseau, France}

\author{Karsten Rode}
\altaffiliation[Now at ]{Physics Department, Trinity College, Dublin 2, Ireland}

\affiliation{Unit\'e Mixte de Physique CNRS/Thales, Route D\'epartementale 128, 91767 Palaiseau, France}

\author{Peter Bencok}
\affiliation{European Synchrotron Radiation Facility (ESRF), 38043 Grenoble Cedex, France}

\author{Agn\`es Barth\'el\'emy}
\affiliation{Unit\'e Mixte de Physique CNRS/Thales, Route D\'epartementale 128, 91767 Palaiseau, France}

\date{\today}

\begin{abstract}


Single phase (001)-oriented BiFeO$_3$ (BFO) thin films grown by pulsed laser deposition can only be obtained in
a narrow window of deposition pressure and temperature and have a low magnetic moment. Out of the stability
window Fe- or Bi-rich impurity phases form, which has a strong impact on the physical and structural properties
of the films, even for impurity concentrations hardly detectable by standard X-ray diffraction measurements. By
using more sensitive tools such as X-Ray absorption spectroscopy and X-ray magnetic circular dichroism and
performing advanced X-ray diffraction characterization, we show that in non-optimal conditions Fe forms
ferrimagnetic $\gamma$-Fe$_2$O$_3$ precipitates that are responsible for virtually all the ferromagnetic signal
measured on such BFO films by standard magnetometry. This confirms that the BFO phase has a very low intrinsic
moment that does not depend on strain. We also study the influence of film thickness on the nucleation of
parasitic phases and find that epitaxial strain can stabilize the pure BFO phase in slightly over-oxidizing
growth conditions.

\end{abstract}

\maketitle

Multiferroic materials display simultaneously several types of order, like ferroelectricity and magnetism
\cite{hill2000,smolenskii82}. Beside their exciting physics, multiferroics could bring solutions for
applications in many fields \cite{wood74}, like agile electromagnetics, optoelectronics or spintronics
\cite{zutic2004}. Indeed, the magnetoelectric coupling existing in multiferroics \cite{smolenskii82,fiebig2005}
could allow to reverse magnetization by applying an electric field \cite{binek2005,binek2005b} instead of a
magnetic field.

To achieve this, a straightforward way would be to use a ferromagnetic layer coupled to a ferromagnetic
multiferroic. However, such ferromagnetic multiferroics are extremely rare, a known exception being BiMnO$_3$
\cite{hill2000} that has a magnetic Curie point (T$_C$) well below room temperature. In fact, most multiferroics
are antiferromagnetic or weak ferromagnets. In this context, the report of a gigantic enhancement of the
magnetic moment in BiFeO$_3$ (BFO) thin films compared to bulk \cite{wang2003} was very exciting given the high
magnetic transition temperature of this compound (640K). Yet, some controversy over these results appeared
shortly after publication and the intrinsic magnetic properties of BiFeO$_3$ films are still debated
\cite{bea2005,eerenstein2005}.

A few months ago, we reported that single-phase BFO films can only be obtained in a rather narrow range of
deposition pressure P$_{O_2}$ and temperature \cite{bea2005}. We found that films not showing indications of
parasitic Fe-rich phases have a low bulk-like magnetic moment ($\sim$0.02 $\mu_B$/Fe) \cite{bai2005} while films
containing Fe oxides display a ferromagnetic behavior. Even though this suggests that in these impure films all
the ferromagnetic signal comes from the Fe oxides, it is not clear whether the BFO phase in such films could not
have some ferromagnetic moment. In this article, we have addressed this issue more quantitatively and
demonstrate that even in Fe-oxide-rich samples the BFO phase has virtually no magnetic moment, irrespective of
its strain state. We also address the influence of strain on the nucleation of extra-phases and find that in
over-oxidizing conditions tending to favour the presence of Bi oxides, strain actually helps to stabilize the
BFO phase.

The films were grown by pulsed laser deposition on (001)-oriented SrTiO$_3$ (STO) substrates \cite{bea2005}. In
order to study the structure of the films, we performed high-resolution X-ray diffraction (XRD) using a
Panalytical X'pert PRO equipped with a Ge (220) monochromator. Magnetization loops were measured at 10K with the
magnetic field oriented in-plane using a Superconducting Quantum Interference Device (SQUID).

X-ray Absorption Spectroscopy (XAS) and X-ray Circular Magnetic Dichroism (XMCD) were performed at the ID08 line
of the European Synchrotron Radiation Facility at the Fe L$_{2,3}$ edges, at 10K. Both magnetic field (in the
range of $\pm$ 6T) and propagation vector of photons were perpendicular to the sample surface with a circular
polarization of nearly 100\%. The spectra were collected in the total electron yield mode that has a typical
probing depth of $\sim$50$\rm{\AA}$ in oxides.

\begin{table} [!h]
 \caption{Composition of the films in the different growth conditions.} \label{tableau}
\begin{ruledtabular}
\begin{tabular}{cccccc}

Film  & ~P$_{O_2}$ (mbar) ~ & $t$ (nm) & ~ \%BFO ~ &  \%$\gamma$-Fe$_2$O$_3$ ~ &  \%Bi$_2$O$_3$ ~ \\
$\sharp$1& 10$^{-4}$ ~ & 120 & ~ 52$\pm$1.5  ~ & ~ 48$\pm$1.5 ~ &  $<$0.4 \\
$\sharp$2& 10$^{-3}$ ~ & 25 & ~ 81$\pm$2  ~ & ~ 19$\pm$2 ~ &  $<$1.7 \\
$\sharp$3& 10$^{-3}$ ~ & 60 & ~ 79$\pm$2  ~ & ~ 21$\pm$2 ~ &  $<$0.7 \\
$\sharp$4& 10$^{-3}$ ~ & 100 & ~ 87$\pm$1.5  ~ & ~ 13$\pm$1.5  & ~ $<$0.5 \\
$\sharp$5& 6 10$^{-3}$ ~ & 35 & $>$97.2  & $<$1.4  &  $<$1.4 \\
$\sharp$6& 1.2 10$^{-2}$ ~ & 70 & $>$98.7   & $<$0.7   &  $<$0.6 \\
$\sharp$7& 1.2 10$^{-2}$ ~ & 120 & 83$\pm$2   & $<$0.4   &  17$\pm$2 \\
$\sharp$8& 1.2 10$^{-2}$ ~ & 240 & 79$\pm$2   & $<$0.2   &  21$\pm$2 \\

\vspace{-1em}

\end{tabular}


\end{ruledtabular}
\end{table}

\begin{figure}[!h]
 \includegraphics[keepaspectratio=true,width=0.9\columnwidth]{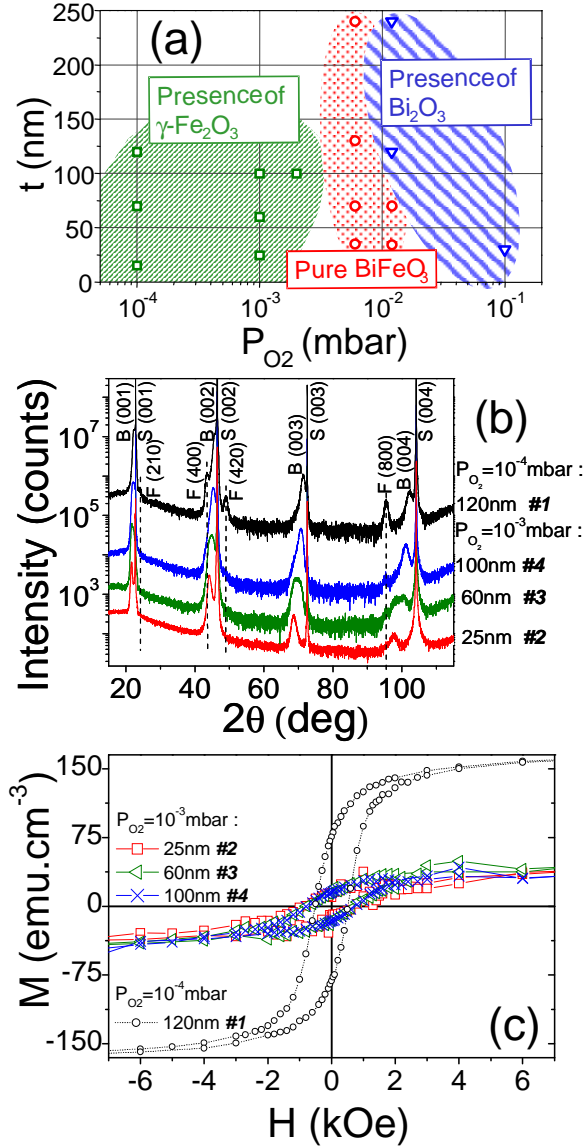}
 \caption{(a) Film composition vs growth pressure and film thickness. (b) XRD diagram of films $\sharp$1 to 4.
 B stands for BFO peaks,
S for STO, and F for the $\gamma$-Fe$_2$O$_3$ parasitic phase. The vertical dashed lines show the location of
the $\gamma$-Fe$_2$O$_3$ reflections. (c) SQUID measurements of the same samples (T=10K)} \label{xrd1}
\end{figure}

For this study, we have fixed the deposition temperature to 580$^\circ$C and varied P$_{O_2}$ and the deposition
time, hence the film thickness (t). We then analysed the influence of pressure and thickness on the onset of
parasitic phases via XRD. The proportions of the different phases have been calculated (table \ref{tableau}) and
the results are summarized in figure \ref{xrd1}a. For P$_{O_2}$ $\leq$ 10$^{-3}$ mbar, Fe oxides are detected in
all the films, irrespective of thickness. The diffraction lines for this Fe oxide correspond to those of
$\gamma$-Fe$_2$O$_3$ (maghemite), a ferrimagnetic material with a T$_C$ of $\sim$850K and a magnetic moment of
$\sim$420 emu.cm$^{-3}$ or 1.25 $\mu_B$/Fe.\cite{hunt95} At P$_{O_2}$=1.2 10$^{-2}$ mbar, we can notice that for
t $\leq$ 70 nm, no parasitic phases are observed while for t $\geq$ 120 nm, Bi$_2$O$_3$ phases are detected.
This suggests different mechanisms for the appearance of Fe- and Bi-rich phases. We will study in the following
the influence of these extra-phases on the structure and properties of the films.

At this point it is important to recall that all films without parasitic phases (e.g. films grown at 6 10$^{-3}$
mbar) have very low magnetic moments ($\leq$0.02 $\mu_B$/Fe) \cite{bea2005}. In figure \ref{xrd1}b, we show XRD
spectra for films $\sharp$1 to 4 grown at 10$^{-4}$ and 10$^{-3}$ mbar. In addition to the (00\emph{l}) peaks
due to STO and BFO, several peaks corresponding to $\gamma$-Fe$_2$O$_3$ reflections are detected. This is
particularly clear for the film $\sharp$1. At 10$^{-3}$ mbar only the (800) peak of $\gamma$-Fe$_2$O$_3$ is
visible but lies on the edge of the BFO (004) peak. This parasitic phase is thus not easily detectable by XRD
and from these data it is not possible to conclude on the presence or absence of $\gamma$-Fe$_2$O$_3$ in the
thinner film. We also note that for the 10$^{-3}$ mbar series, the position of the BFO peaks varies due to
strain relaxation occurring upon increasing t.

Figure \ref{xrd1}c shows magnetization hysteresis cycles M(H) for these four films. Remarkably, the three
samples grown at P$_{O_2}$=10$^{-3}$ mbar show the same volumic magnetic moment. For the film $\sharp$1, the
volumic magnetic moment is larger which also corresponds to a larger proportion of $\gamma$-Fe$_2$O$_3$ as
detected by XRD.

\begin{figure}[!h]
 \includegraphics[width=0.9\columnwidth]{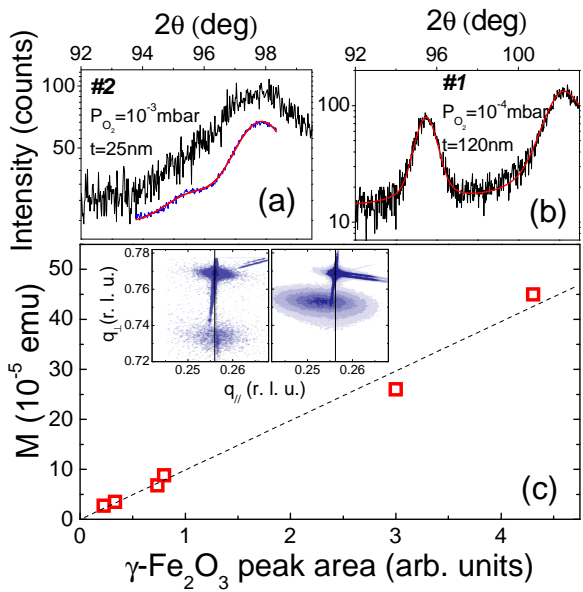}
 \caption{(a) XRD pattern in the 92-100$^\circ$ 2$\theta$-range of film $\sharp$2. The upper spectrum
is a zoom of the fig \ref{xrd1}b. The lower one is a pattern measured with a higher counting rate (see text).
The bold line is a fit of the (004) BFO and (800) $\gamma$-Fe$_2$O$_3$ peaks. (b) Zoom of the XRD pattern of
film $\sharp$1 shown in figure \ref{xrd1}b. The bold line is a fit of the BFO and $\gamma$-Fe$_2$O$_3$ peaks.
(c) Saturation magnetization for samples grown at 10$^{-3}$ and 10$^{-4}$ mbar as a function of the area of the
(800) $\gamma$-Fe$_2$O$_3$ peak. The reciprocal space maps of the (013) reflections reveal a fully strained
state for a 25 nm film (left) and a partially relaxed state for a 100 nm film (right), both grown at 10$^{-3}$
mbar. r. l. u. stands for reciprocal lattice units.} \label{xrd2}
\end{figure}

The observation of a large magnetic moment for the film $\sharp$2 suggests the presence of $\gamma$-Fe$_2$O$_3$
even though standard XRD could not detect it. To solve this issue we performed additional XRD with much longer
counting times in the 2$\theta$ = 93-98$^\circ$ range so as to increase the signal-to-noise ratio, see figure
\ref{xrd2}a. On this scan, a shoulder is clearly visible at the left of the BFO (004) peak, located at 2$\theta
\approx 95.5 ^\circ$ , i.e. where the (800) reflection of $\gamma$-Fe$_2$O$_3$ shows up for the thicker films.

To better quantify the relation between the amount of maghemite and the magnetization, we have fitted the XRD
spectra in the 2$\theta$ = 90-98$^\circ$ range with three pseudo-Voigt functions corresponding to the (004) BFO
peak, the (004) STO peak and the (800) $\gamma$-Fe$_2$O$_3$ peak. The STO fit allows to normalize the area of
the $\gamma$-Fe$_2$O$_3$ peak to that of a substrate peak. We thus obtain the area A of the $\gamma$-Fe$_2$O$_3$
(800) peak, proportional to the volume of $\gamma$-Fe$_2$O$_3$ present in the film (see an example of such fits
in figure \ref{xrd2}b). As visible in figure \ref{xrd2}c, M increases linearly with A, which shows that the
magnetic signal in these films is proportional to the amount of $\gamma$-Fe$_2$O$_3$. Of great relevance is the
observation that the magnetization extrapolates to zero for A=0 hence demonstrating that in these films,
virtually \emph{all} the magnetic signal comes from $\gamma$-Fe$_2$O$_3$. In other words, the BFO phase has a
very low magnetization, if any. Reciprocal space mappings collected on the films considered in the above
analysis reveal a variety of strain states (see insets in fig \ref{xrd2}c). Hence, our observation of a very low
moment for the BFO phase in all these films rules out a role of strain in increasing the BFO magnetization for
this range of growth pressure and temperature. This is in contrast to what was suggested by Wang \emph{et al}
\cite{wang2003} for higher growth pressure and temperature but in agreement with theoretical predictions
\cite{ederer2005} and with Eerenstein \emph{et al} \cite{eerenstein2005}.

\begin{figure}[!h]
 \includegraphics[keepaspectratio=true,width=0.9\columnwidth]{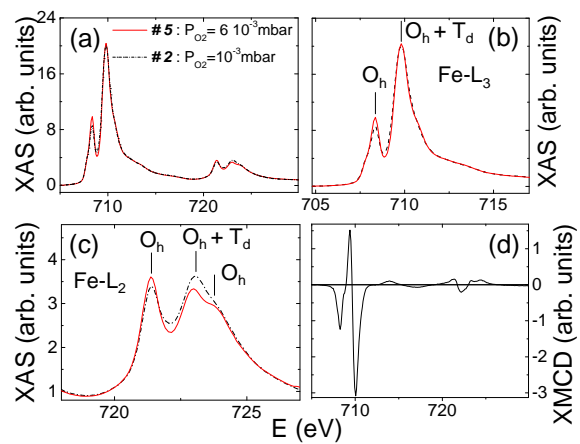}
 \caption{(a) XAS spectra at the Fe L$_{2,3}$ edge for films $\sharp$2 and 5. (b) and (c) are zooms of (a) next
 to the L$_3$ and L$_2$ edges respectively. The lines
correspond to an attribution of the Fe$^{3+}$ peaks from multiplets calculation \protect\cite{profeta2005}. (d)
XMCD of the sample $\sharp$2.} \label{xas}
\end{figure}

To get deeper insight on the microscopic magnetic properties of our BFO films, we have performed XAS and XMCD
measurements on films $\sharp$2 and $\sharp$5. XAS at the L$_{2,3}$ edge of Fe allows the determination of its
valence and environment. In pure BFO, Fe is octahedrally coordinated by 6 oxygen ions and has a valence of 3+.
In $\gamma$-Fe$_2$O$_3$ the iron valence is also 3+ but 5 of the 8 Fe ions contained in a unit-cell are in
octahedral (Oh) sites while the other 3 are in tetrahedral (Td) sites \cite{coey71}. In figure \ref{xas}a, we
show XAS spectra for both films. The general shape of the spectra is very similar to that obtained on LaFeO$_3$,
a perovskite in which Fe$^{3+}$ ions are in a octahedral environment \cite{czekaj2006}. A large difference is
observed with a XAS spectrum for Fe$_3$O$_4$ \cite{crocombette95,morrall2003} in which 1/3 of the iron ions are
in a 2+ state, strongly suggesting that Fe ions have a valence close to 3+ in both films.

A closer look at the XAS data (figure \ref{xas}b and c) allows the identification of several differences between
the two films. Especially, the shoulder at the right of the 723 eV peak of the L$_2$ edge, typical of Fe$^{3+}$
in an octahedral environment \cite{profeta2005}, is clearly visible for film $\sharp$5 but less pronounced for
film $\sharp$2. This indicates a substantial amount of Fe$^{3+}$ at Td sites in this latter sample. Consistent
with this observation is the lower XAS signal measured at the 721.5 eV peak. We conclude that film $\sharp$2 has
a larger proportion of Fe$^{3+}$ in Td sites, as expected in the presence of $\gamma$-Fe$_2$O$_3$.

In film $\sharp$5 no clear XMCD signal was measured, suggesting a very low magnetic moment, in agreement with
SQUID results. On sample $\sharp$2 we find a large dichroic signal (see figure \ref{xas}d), very similar to that
of $\gamma$-Fe$_2$O$_3$ \cite{profeta2005,crocombette95}. More quantitatively, we can calculate the spin moment
M$_S$ and the orbital moment M$_L$ using the sum rules, which yields M$_S=0.22 \pm 0.05 \mu_B$/Fe and M$_L=0.02
\pm 0.005 \mu_B$/Fe. The total magnetic moment is M = M$_S$+M$_L= 0.24 \pm 0.06 \mu_B$/Fe, in very good
agreement with the magnetization measured by SQUID, 40 emu.cm$^{-3}$ or 0.27 $\mu_B$/Fe. Since bulk maghemite
has a magnetic moment of 1.25 $\mu_B$/Fe, the proportion of $\gamma$-Fe$_2$O$_3$ near the surface (probed by
XMCD) is thus close to the 20\% calculated for the volume (see table \ref{tableau}). Therefore, the distribution
of this extra phase in the film seems to be homogeneous, which is consistent with the presence of
$\gamma$-Fe$_2$O$_3$ for all thicknesses, particularly for the lower ones.

\begin{figure}[!h]
 \includegraphics[keepaspectratio=true,width=0.9\columnwidth]{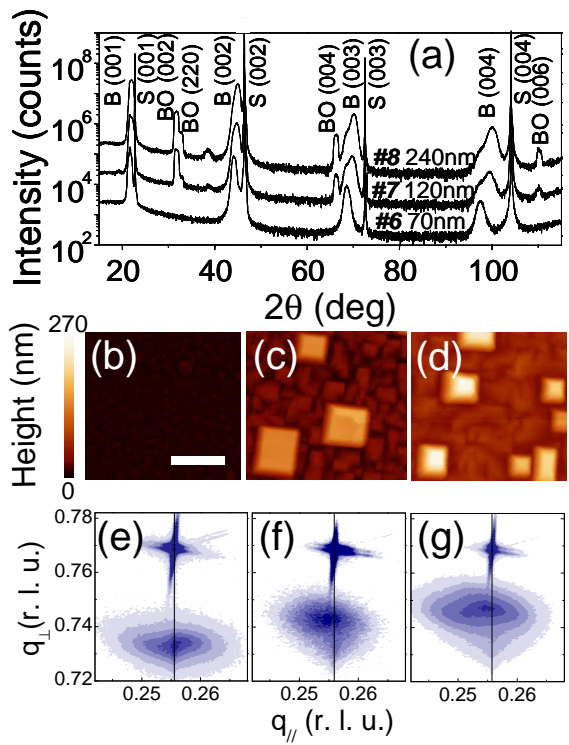}
 \caption{(a) XRD pattern of samples $\sharp$6, 7 and 8. The peaks
are indexed as B for BFO, S for STO and BO for the extra-phase Bi$_2$O$_3$. AFM [(b), (c), (d)] and reciprocal
space mappings around the (013) direction [(e), (f), (g)] for films $\sharp$6 [(b) and (e)], $\sharp$7 [(c) and
(f)] and $\sharp$8 [(d) and (g)]. For the AFM images, the white bar corresponds to 1 $\mu$m.} \label{bo}
\end{figure}

We now examine in more detail the high-pressure region of the diagram shown in figure \ref{xrd1}a. In figure
\ref{bo}a, XRD spectra of samples grown at 1.2 10$^{-2}$ mbar with different thicknesses are shown. For the
thinner film, no parasitic phases are observed, even for long counting rates (not shown), while at larger t
Bi$_2$O$_3$ lines show up. On the AFM images of fig \ref{bo}b-d, we can see that the detection of Bi$_2$O$_3$ is
accompanied by the nucleation of square outgrowths on the surface for t $\geq$120 nm. Auger electron
spectroscopy confirmed that these hillocks are Bi-rich and Fe-poor, thus likely corresponding to Bi$_2$O$_3$
(Ref \onlinecite{bea2005}). When t increases, these outgrowths become higher (as high as the film thickness).

It is interesting to correlate the nucleation of Bi$_2$O$_3$ with the strain state of the films using reciprocal
space maps close to the (013) reflection, see figures \ref{bo}e-g. The relative positions of the STO and BFO
peaks indicate a fully strained state for the 70 nm film while they correspond to a partially relaxed state for
the 120 and 240 nm films. This correlation between strain relaxation and the nucleation of Bi$_2$O$_3$ suggests
that strain favours the formation of a metastable BFO phase at the expense of the stable Bi$_2$O$_3$. The
stabilization of metastable perovskite phases by strain has indeed been reported by several groups
\cite{novojilov2000,samoylenkov99}. More quantitatively, the formation energy difference $\Delta E$ between the
stable and the metastable phase is given by:

\begin{equation}
\Delta E =t\Big[(\Delta g_S - \Delta g_M) - \frac{\mu}{1-\nu}\epsilon^2 \Big] + (\sigma^I_S - \sigma^I_M)
\label{eq}
\end{equation}

\noindent $\Delta g_S$ and $\Delta g_M$ are the stable and metastable phase formation energies, respectively,
$\mu$ and $\nu$ the shear and Poisson coefficients of the metastable phase, $\epsilon$ the strain and
$\sigma^I_S$ and $\sigma^I_M$ the interface energies for the stable and metastable phases, respectively. In
principle the first term is always negative. The second term is positive if, for instance, the interface between
the metastable phase and the substrate is coherent (i.e. the metastable phases grows in a fully strained state)
while the interface between the stable phase and the substrate is not (e.g. if the mismatch between the stable
phase and the substrate is large). Thus, when t is small $\Delta E$ can be positive and the growth of the
metastable phase is favoured but there is a critical thickness t$_c$ above which $\Delta E$ becomes negative
which leads to the formation of the stable phase. On the other hand, if the structure of the metastable phase
relaxes (hence enhancing its interface energy) the formation of the stable phase can become more favourable for
t $<$ t$_c$.

In our case, we have seen that BFO grows fully strained on the STO substrate up to t $\approx$ 70 nm so that the
BFO/STO interface energy $\sigma_{BFO/STO}$ is low and the conditions for the growth of the metastable BFO phase
are met. The formation of Bi$_2$O$_3$ occurs when BFO relaxes, hence increasing $\sigma_{BFO/STO}$. It is thus
tempting to conclude that this relaxation unbalances equation \ref{eq} and favours the formation of the stable
Bi$_2$O$_3$ phase. However, since many parameters in equation \ref{eq} are unknown, it is not possible to
calculate t$_c$. Thus we cannot be fully conclusive on whether the formation of Bi$_2$O$_3$ is just related to
the thickness increase or directly due to the observed structural relaxation.

In summary, we have found that epitaxial BFO films grown at 6 10$^{-3}$ mbar are single-phase for thicknesses up
to at least 240 nm. For lower pressures, all the films contain $\gamma$-Fe$_2$O$_3$, as evidenced by XRD, XAS
and XMCD. We showed that this extra phase is responsible for all the magnetic moment in the film so that, at
least in all the range of growth conditions we have explored, the BFO phase has a very low magnetic moment, if
any. At 1.2 10$^{-2}$ mbar BFO films are also single-phase for t$\leq$70 nm while Bi$_2$O$_3$ appears in the
film above that thickness. Reciprocal space maps analysis suggests a role of strain in stabilizing a metastable
BFO phase at this pressure.

\acknowledgements{Enlightening discussions with F. Petroff, V. Cros and K. Bouzehouane are gratefully
acknowledged. This work has been supported by the E.U. STREP MACOMUFI (033221) and the contract FEMMES of the
Agence Nationale pour la Recherche. H.B. also acknowledges financial support from the Conseil G\'en\'eral de
l'Essonne.}


\begin{thebibliography}{21}
\expandafter\ifx\csname natexlab\endcsname\relax\def\natexlab#1{#1}\fi \expandafter\ifx\csname
bibnamefont\endcsname\relax
  \def\bibnamefont#1{#1}\fi
\expandafter\ifx\csname bibfnamefont\endcsname\relax
  \def\bibfnamefont#1{#1}\fi
\expandafter\ifx\csname citenamefont\endcsname\relax
  \def\citenamefont#1{#1}\fi
\expandafter\ifx\csname url\endcsname\relax
  \def\url#1{\texttt{#1}}\fi
\expandafter\ifx\csname urlprefix\endcsname\relax\def\urlprefix{URL }\fi \providecommand{\bibinfo}[2]{#2}
\providecommand{\eprint}[2][]{\url{#2}}

\bibitem[{\citenamefont{{N.A. Hill}}(2000)}]{hill2000}
\bibinfo{author}{\bibnamefont{{N.A. Hill}}}, \bibinfo{journal}{J. Phys. Chem.
  B} \textbf{\bibinfo{volume}{104}}, \bibinfo{pages}{6694}
  (\bibinfo{year}{2000}).

\bibitem[{\citenamefont{{G.A. Smolenskii} and {I.E.
  Chupis}}(1982)}]{smolenskii82}
\bibinfo{author}{\bibnamefont{{G.A. Smolenskii}}} \bibnamefont{and}
  \bibinfo{author}{\bibnamefont{{I.E. Chupis}}}, \bibinfo{journal}{Sov. Phys.
  Usp.} \textbf{\bibinfo{volume}{25}}, \bibinfo{pages}{475}
  (\bibinfo{year}{1982}).

\bibitem[{\citenamefont{{V.E. Wood} and {A.E. Austin}}(1974)}]{wood74}
\bibinfo{author}{\bibnamefont{{V.E. Wood}}} \bibnamefont{and}
  \bibinfo{author}{\bibnamefont{{A.E. Austin}}}, \bibinfo{journal}{Int. J.
  Magn.} \textbf{\bibinfo{volume}{5}}, \bibinfo{pages}{303}
  (\bibinfo{year}{1974}).

\bibitem[{\citenamefont{\v{Z}uti\'{c} et~al.}(2004)\citenamefont{\v{Z}uti\'{c},
  Fabian, and {Das Sarma}}}]{zutic2004}
\bibinfo{author}{\bibfnamefont{I.}~\bibnamefont{\v{Z}uti\'{c}}},
  \bibinfo{author}{\bibfnamefont{J.}~\bibnamefont{Fabian}}, \bibnamefont{and}
  \bibinfo{author}{\bibfnamefont{S.}~\bibnamefont{{Das Sarma}}},
  \bibinfo{journal}{Rev. Mod. Phys.} \textbf{\bibinfo{volume}{76}},
  \bibinfo{pages}{323} (\bibinfo{year}{2004}).

\bibitem[{\citenamefont{Fiebig}(2005)}]{fiebig2005}
\bibinfo{author}{\bibfnamefont{M.}~\bibnamefont{Fiebig}}, \bibinfo{journal}{J.
  Phys. D.: Appl. Phys.} \textbf{\bibinfo{volume}{38}}, \bibinfo{pages}{R123}
  (\bibinfo{year}{2005}).

\bibitem[{\citenamefont{Binek and Doudin}(2005)}]{binek2005}
\bibinfo{author}{\bibfnamefont{Ch.}~\bibnamefont{Binek}} \bibnamefont{and}
  \bibinfo{author}{\bibfnamefont{B.}~\bibnamefont{Doudin}},
  \bibinfo{journal}{J. Phys.: Condens. Matter} \textbf{\bibinfo{volume}{17}},
  \bibinfo{pages}{L39} (\bibinfo{year}{2005}).

\bibitem[{\citenamefont{Binek et~al.}(2005)}]{binek2005b}
\bibinfo{author}{\bibfnamefont{Ch.}~\bibnamefont{Binek}},
    \bibinfo{author}{\bibfnamefont{A.}~\bibnamefont{Hochstrat}},
    \bibinfo{author}{\bibfnamefont{X.}~\bibnamefont{Chen}},
    \bibinfo{author}{\bibfnamefont{P.}~\bibnamefont{Borisov}},
    \bibinfo{author}{\bibfnamefont{W.}~\bibnamefont{Kleemann}} \bibnamefont{and}
    \bibinfo{author}{\bibfnamefont{B.}~\bibnamefont{Doudin}},
  \bibinfo{journal}{J. Appl. Phys.} \textbf{\bibinfo{volume}{97}},
  \bibinfo{pages}{10C514} (\bibinfo{year}{2005}).

\bibitem[{\citenamefont{Wang et~al.}(2003)}]{wang2003}
\bibinfo{author}{\bibfnamefont{J.}~\bibnamefont{Wang}},
    \bibinfo{author}{\bibfnamefont{J.B.}~\bibnamefont{Neaton}},
    \bibinfo{author}{\bibfnamefont{H.}~\bibnamefont{Zheng}},
    \bibinfo{author}{\bibfnamefont{V.}~\bibnamefont{Nagarajan}},
    \bibinfo{author}{\bibfnamefont{S.B.}~\bibnamefont{Ogale}},
    \bibinfo{author}{\bibfnamefont{B.}~\bibnamefont{Liu}},
    \bibinfo{author}{\bibfnamefont{D.}~\bibnamefont{Viehland}},
    \bibinfo{author}{\bibfnamefont{V.}~\bibnamefont{Vaithyanathan}},
    \bibinfo{author}{\bibfnamefont{D.G.}~\bibnamefont{Schlom}},
    \bibinfo{author}{\bibfnamefont{U.V.}~\bibnamefont{Waghmare}},
    \bibinfo{author}{\bibfnamefont{N.A.}~\bibnamefont{Spaldin}},
    \bibinfo{author}{\bibfnamefont{K.M.}~\bibnamefont{Rabe}},
    \bibinfo{author}{\bibfnamefont{M.}~\bibnamefont{Wuttig}} \bibnamefont{and}
    \bibinfo{author}{\bibfnamefont{R.}~\bibnamefont{Ramesh}},
  \bibinfo{journal}{Science} \textbf{\bibinfo{volume}{299}},
  \bibinfo{pages}{1719} (\bibinfo{year}{2003}).

\bibitem[{\citenamefont{B{\'e}a et~al.}(2005)}]{bea2005}
\bibinfo{author}{\bibfnamefont{H.}~\bibnamefont{B{\'e}a}},
    \bibinfo{author}{\bibfnamefont{M.}~\bibnamefont{Bibes}},
    \bibinfo{author}{\bibfnamefont{A.}~\bibnamefont{Barth{\'e}l{\'e}my}},
    \bibinfo{author}{\bibfnamefont{K.}~\bibnamefont{Bouzehouane}},
    \bibinfo{author}{\bibfnamefont{A.}~\bibnamefont{Khodan}},
    \bibinfo{author}{\bibfnamefont{J.-P.}~\bibnamefont{Contour}},
    \bibinfo{author}{\bibfnamefont{S.}~\bibnamefont{Fusil}},
    \bibinfo{author}{\bibfnamefont{F.}~\bibnamefont{Wyczisk}},
    \bibinfo{author}{\bibfnamefont{A.}~\bibnamefont{Forget}},
    \bibinfo{author}{\bibfnamefont{D.}~\bibnamefont{Lebeugle}},
    \bibinfo{author}{\bibfnamefont{D.}~\bibnamefont{Colson}} \bibnamefont{and}
    \bibinfo{author}{\bibfnamefont{M.}~\bibnamefont{Viret}},
  \bibinfo{journal}{Appl. Phys. Lett.} \textbf{\bibinfo{volume}{88}},
  \bibinfo{pages}{062502} (\bibinfo{year}{2005}).

\bibitem[{\citenamefont{Eerenstein et~al.}(2005)}]{eerenstein2005}
\bibinfo{author}{\bibfnamefont{W.}~\bibnamefont{Eerenstein}},
    \bibinfo{author}{\bibfnamefont{F.D.}~\bibnamefont{Morrison}},
  \bibinfo{author}{\bibfnamefont{J.}~\bibnamefont{Dho}},
    \bibinfo{author}{\bibfnamefont{M.G.}~\bibnamefont{Blamire}},
    \bibinfo{author}{\bibfnamefont{J.F.}~\bibnamefont{Scott}} \bibnamefont{and}
    \bibinfo{author}{\bibfnamefont{N.D.}~\bibnamefont{Mathur}},
  \bibinfo{journal}{Science}
  \textbf{\bibinfo{volume}{307}}, \bibinfo{pages}{1203a}
  (\bibinfo{year}{2005}).

\bibitem[{\citenamefont{Bai et~al.}(2005)}]{bai2005}
\bibinfo{author}{\bibfnamefont{F.}~\bibnamefont{Bai}},
    \bibinfo{author}{\bibfnamefont{J.}~\bibnamefont{Wang}},
    \bibinfo{author}{\bibfnamefont{M.}~\bibnamefont{Wuttig}},
    \bibinfo{author}{\bibfnamefont{J.}~\bibnamefont{Li}},
    \bibinfo{author}{\bibfnamefont{N.}~\bibnamefont{Wang}},
    \bibinfo{author}{\bibfnamefont{A.P.}~\bibnamefont{Pyatakov}},
    \bibinfo{author}{\bibfnamefont{A.K.}~\bibnamefont{Zvezdin}},
    \bibinfo{author}{\bibfnamefont{L.E.}~\bibnamefont{Cross}} \bibnamefont{and}
    \bibinfo{author}{\bibfnamefont{D.}~\bibnamefont{Viehland}},
  \bibinfo{journal}{Appl. Phys. Lett.} \textbf{\bibinfo{volume}{86}},
  \bibinfo{pages}{032511} (\bibinfo{year}{2005}).

\bibitem[{\citenamefont{{C.P. Hunt} et~al.}(1995)\citenamefont{{C.P. Hunt},
  {B.M. Moskowitz}, and {S.K. Banerjee}}}]{hunt95}
\bibinfo{author}{\bibnamefont{{C.P. Hunt}}},
  \bibinfo{author}{\bibnamefont{{B.M. Moskowitz}}}, \bibnamefont{and}
  \bibinfo{author}{\bibnamefont{{S.K. Banerjee}}}, \emph{\bibinfo{title}{Rock
  Physics and Phase Transitions, a Handbook of Physical Constants}}
  (\bibinfo{year}{1995}).

\bibitem[{\citenamefont{Ederer and {N.A. Spaldin}}(2005)}]{ederer2005}
\bibinfo{author}{\bibfnamefont{C.}~\bibnamefont{Ederer}} \bibnamefont{and}
  \bibinfo{author}{\bibnamefont{{N.A. Spaldin}}}, \bibinfo{journal}{Phys. Rev.
  B} \textbf{\bibinfo{volume}{71}}, \bibinfo{pages}{224103}
  (\bibinfo{year}{2005}).

\bibitem[{\citenamefont{{J.M.D. Coey}}(1971)}]{coey71}
\bibinfo{author}{\bibnamefont{{J.M.D. Coey}}}, \bibinfo{journal}{Phys. Rev.
  Lett.} \textbf{\bibinfo{volume}{27}}, \bibinfo{pages}{1140}
  (\bibinfo{year}{1971}).

\bibitem[{\citenamefont{Czekaj et~al.}(2006)\citenamefont{Czekaj, Nolting,
  {L.J. Heyderman}, {P.R. Wilmott}, and {G. van der Laan}}}]{czekaj2006}
\bibinfo{author}{\bibfnamefont{S.}~\bibnamefont{Czekaj}},
  \bibinfo{author}{\bibfnamefont{F.}~\bibnamefont{Nolting}},
  \bibinfo{author}{\bibnamefont{{L.J. Heyderman}}},
  \bibinfo{author}{\bibnamefont{{P.R. Willmott}}}, \bibnamefont{and}
  \bibinfo{author}{\bibnamefont{{G. van der Laan}}}, \bibinfo{journal}{Phys.
  Rev. B} \textbf{\bibinfo{volume}{73}}, \bibinfo{pages}{020401(R)}
  (\bibinfo{year}{2006}).

\bibitem[{\citenamefont{Morrall et~al.}(2003)}]{morrall2003}
\bibinfo{author}{\bibfnamefont{P.}~\bibnamefont{Morrall}},
    \bibinfo{author}{\bibfnamefont{F.}~\bibnamefont{Schedin}},
    \bibinfo{author}{\bibfnamefont{G.S.}~\bibnamefont{Case}},
    \bibinfo{author}{\bibfnamefont{M.F.}~\bibnamefont{Thomas}},
    \bibinfo{author}{\bibfnamefont{E.}~\bibnamefont{Dudzik}},
    \bibinfo{author}{\bibfnamefont{G.}~\bibnamefont{van der Laan}} \bibnamefont{and}
    \bibinfo{author}{\bibfnamefont{G.}~\bibnamefont{Thornton}},
  \bibinfo{journal}{Phys. Rev. B} \textbf{\bibinfo{volume}{67}},
  \bibinfo{pages}{214408} (\bibinfo{year}{2003}).

\bibitem[{\citenamefont{{J.P. Crocombette} et~al.}(1995)\citenamefont{{J.P.
  Crocombette}, Pollak, Jollet, Thromat, and {M.
  Gautier-Soyer}}}]{crocombette95}
\bibinfo{author}{\bibnamefont{{J.P. Crocombette}}},
  \bibinfo{author}{\bibfnamefont{M.}~\bibnamefont{Pollak}},
  \bibinfo{author}{\bibfnamefont{F.}~\bibnamefont{Jollet}},
  \bibinfo{author}{\bibfnamefont{N.}~\bibnamefont{Thromat}}, \bibnamefont{and}
  \bibinfo{author}{\bibnamefont{{M. Gautier-Soyer}}}, \bibinfo{journal}{Phys.
  Rev. B} \textbf{\bibinfo{volume}{52}}, \bibinfo{pages}{3143}
  (\bibinfo{year}{1995}).

\bibitem[{\citenamefont{{S. Brice-Profeta} et~al.}(2005)}]{profeta2005}
\bibinfo{author}{\bibnamefont{{S. Brice-Profeta}}},
    \bibinfo{author}{\bibfnamefont{M.-A.}~\bibnamefont{Arrio}},
    \bibinfo{author}{\bibfnamefont{E.}~\bibnamefont{Tronc}},
    \bibinfo{author}{\bibfnamefont{N.}~\bibnamefont{Menguy}},
    \bibinfo{author}{\bibfnamefont{I.}~\bibnamefont{Letard}},
    \bibinfo{author}{\bibfnamefont{C.}~\bibnamefont{Cartier dit Moulin}},
    \bibinfo{author}{\bibfnamefont{M.}~\bibnamefont{Nogu{\`e}s}},
    \bibinfo{author}{\bibfnamefont{C.}~\bibnamefont{Chan{\'e}ac}},
    \bibinfo{author}{\bibfnamefont{J.-P.}~\bibnamefont{Jolivet}} \bibnamefont{and}
    \bibinfo{author}{\bibfnamefont{Ph.}~\bibnamefont{Sainctavit}},
  \bibinfo{journal}{J. Magn. Magn. Mater.} \textbf{\bibinfo{volume}{288}},
  \bibinfo{pages}{354} (\bibinfo{year}{2005}).

\bibitem[{\citenamefont{{M.A. Novojilov} et~al.}(2000)}]{novojilov2000}
\bibinfo{author}{\bibnamefont{{M.A. Novojilov}}},
    \bibinfo{author}{\bibfnamefont{O.Y.}~\bibnamefont{Gorbenko}},
    \bibinfo{author}{\bibfnamefont{I.E.}~\bibnamefont{Graboy}},
    \bibinfo{author}{\bibfnamefont{A.R.}~\bibnamefont{Kaul}},
    \bibinfo{author}{\bibfnamefont{H.W.}~\bibnamefont{Zandbergen}},
    \bibinfo{author}{\bibfnamefont{N.A.}~\bibnamefont{Babushkina}} \bibnamefont{and}
    \bibinfo{author}{\bibfnamefont{L.M.}~\bibnamefont{Belova}},
  \bibinfo{journal}{Appl. Phys. Lett.} \textbf{\bibinfo{volume}{76}},
  \bibinfo{pages}{2041} (\bibinfo{year}{2000}).

\bibitem[{\citenamefont{{S.V. Samoylenkov} et~al.}(1999)}]{samoylenkov99}
\bibinfo{author}{\bibnamefont{{S.V. Samoylenkov}}}
    \bibinfo{author}{\bibfnamefont{O.Y.}~\bibnamefont{Gorbenko}},
    \bibinfo{author}{\bibfnamefont{I.E.}~\bibnamefont{Graboy}},
    \bibinfo{author}{\bibfnamefont{A.R.}~\bibnamefont{Kaul}},
    \bibinfo{author}{\bibfnamefont{H.W.}~\bibnamefont{Zandbergen}} \bibnamefont{and}
    \bibinfo{author}{\bibfnamefont{E.}~\bibnamefont{Connolly}},
  \bibinfo{journal}{Chem. Mater.} \textbf{\bibinfo{volume}{11}},
  \bibinfo{pages}{2417} (\bibinfo{year}{1999}).

\end{thebibliography}
 \end{document}